\documentclass[epsfig,graphics,twocolumn,floatfix,a4paper,showpacs]{revtex4}
\usepackage{amsmath,amsfonts,amssymb,graphicx,color,times,bbm,psfrag,dsfont}
\usepackage[english]{babel}
\usepackage[latin1]{inputenc}
\usepackage[T1]{fontenc}

\newcommand{\beq}{\begin{equation}}
\newcommand{\eeq}{\end{equation}}
\newcommand{\beqa}{\begin{eqnarray}}
\newcommand{\eeqa}{\end{eqnarray}}
\newcommand{\ket} [1] {\vert #1 \rangle}
\newcommand{\bra} [1] {\langle #1 \vert}

\newcommand{\PsiGS}{\Psi_{\mbox{\tiny GS}}}

\usepackage{graphicx}
\usepackage[mathscr]{eucal}
\usepackage{amsmath}
\usepackage{amssymb}
\usepackage{epstopdf}
\usepackage{epsfig}
\def\one{\ensuremath{\hbox{$\mathrm I$\kern-.6em$\mathrm 1$}}}

\begin{document} 

\title{Entanglement growth and simulation efficiency in one-dimensional quantum lattice systems}
\author{\'Alvaro Perales$^{1,2}$ and Guifr\'e Vidal$^2$}

\affiliation{$^1$Dpto.\ Autom\'atica, Universidad de Alcal\'a, Alcal\'a de Henares, Madrid 28871, Spain.\\
$^2$School of Physical Sciences, The University of Queensland, QLD 4072, Australia.}

\begin{abstract}
We study the evolution of one-dimensional quantum lattice systems when the ground state is perturbed by altering one site in the middle of the chain. For a large class of models, we observe a similar pattern of entanglement growth during the evolution, characterized by a moderate increase of significant Schmidt coefficients in all relevant bipartite decompositions of the state. As a result, the evolution can be accurately described by a \emph{matrix product state} and efficiently simulated using the \emph{time-evolving block decimation} algorithm. 
\end{abstract}
\pacs{}
\maketitle

\section{INTRODUCTION}


The numerical study of quantum many-body systems is a challenging computational task due to the exponential growth of the Hilbert space dimension with the system's size. Lattice systems in one spatial dimension, such as quantum spin chains, are a noticeable exception. There, the \emph{density matrix renormalization group} (DMRG) algorithm allows for the precise computation of ground state properties \cite{DMRG}, while the \emph{time-evolving block decimation} (TEBD) algorithm can be used to simulate time evolution \cite{TEBD1,TEBD2}. Such techniques provide valuable insight into quantum systems, therefore facilitating progress in several forefront areas of research, both in science ---e.g. condensed matter, quantum optics, atomic and nuclear physics, quantum chemistry--- and technology ---e.g. quantum information processing, quantum computation, nano-technology.


Many-body entanglement is at the very core of the achievements of the DMRG and TEBD algorithms. In both cases, a key ingredient is the use of a \emph{matrix product state} (MPS) to represent the state $\ket{\Psi}$ of the system \cite{MPS}. A MPS leads to an \emph{efficient} representation of $\ket{\Psi}$ provided that the amount of entanglement in the system is sufficiently small. Thus, the success of the DMRG relies on the remarkable fact that in one spatial dimension the ground state of most local Hamiltonians has only a limited amount of entanglement. Likewise, the TEBD algorithm, based on updating the MPS in time, is efficient as long as no large amounts of entanglement are produced during the simulated evolution.


Identifying time evolutions that only involve small amounts of entanglement is, consequently, of central importance in order to establish the range of applicability of the TEBD algorithm --- and of a whole score of subsequent proposals \cite{TEBD3,White,Daley,tDMRG}, including implementations within the DMRG formalism \cite{White,Daley,tDMRG} often referred to as tDMRG, that are based on the same idea, namely on adapting the MPS representation of the state of the system, so as to account for its changes during a time evolution. At the same time, a better understanding of the dynamics of entanglement in one-dimensional many-body systems \cite{EntDynamics} is of interest in the areas of quantum information processing, where entanglement is regarded as a useful resource, and of quantum computation, where entanglement is necessary in order to achieve a significant speed-up with respect to classical computers \cite{JL,TEBD1}.


In this paper we study the generation of entanglement in a particularly relevant class of time evolutions. Specifically, we consider a chain of $N$ sites, where each site is represented by a finite-dimensional Hilbert space and labeled by $s$, ($s=1,\cdots,N$). The system, initially prepared in the ground state $\ket{\PsiGS}$ of some local Hamiltonian $H$, is perturbed at time $t=0$ by applying an operator $A$ on lattice site $s_0$. As a result, the state of the system $\ket{\Psi(0)} \equiv A^{s_0}\ket{\PsiGS}$, no longer an eigenstate of $H$, evolves non-trivially in time, see e.g. Fig.\ (\ref{HBMottnument3D}),
\begin{equation}
	\ket{\Psi(t)} \equiv e^{-iHt}\ket{\Psi(0)}= e^{-iHt} A^{s_0}\ket{\PsiGS}.
\label{eq:evolution}
\end{equation}
During this evolution entanglement is produced, adding to the entanglement that might already be present in the ground state.


Notice that state $\ket{\Psi(t)}$ appears in the computation of the unequal time correlation function
\begin{eqnarray}
	\left\langle B^s(t)A^{s_0}(0)\right\rangle &\equiv& \bra{\PsiGS} e^{iHt} B^s e^{-iHt} A^{s_0}\ket{\PsiGS} \nonumber\\
	&=& e^{iE_0t}\bra{\PsiGS}B^s\ket{\Psi(t)},
	\label{eq:correlator}
\end{eqnarray}
where a second operator $B$ has been applied on site $s$ at time $t$. Therefore, by characterizing the growth of entanglement in evolution (\ref{eq:evolution}), we will be able to assess the efficiency with which the algorithms of Refs.\ \cite{TEBD1,TEBD2,TEBD3,White,Daley,tDMRG} can be used to compute the two-point correlator (\ref{eq:correlator}). We recall that from this correlator one can extract quantities such as Green's functions or dynamic structure factors, and thus learn about a number of properties of the system, including its response to external probes, e.g. neutron or photon scattering.


The study of the entanglement and simulatability of a time evolution of the form (\ref{eq:evolution}) was initiated in Ref. \cite{TEBD2} and is intimatedly related to the development of the TEBD algorithm. Originally, the TEBD algorithm was created in order to characterize the role of entanglement in quantum computation \cite{TEBD1}. Soon afterward it was noticed that, for systems in one spatial dimension, low energy time evolutions such as (\ref{eq:evolution}) seemed to often involve small amounts of entanglement \cite{TEBD2}, precisely in the way that would allow the TEBD algorithm to work efficiently. This opened up the possibility to simulate this class of dynamics. One may claim, a posteriori, that it is somehow expected that no much entanglement will be created when a single local perturbation is introduced on the ground state $\ket{\PsiGS}$ of $H$. However, the subject is far from being well understood and, for instance, given $H$ and local operator $A$, there are no known conditions to guarantee that the evolution can be simulated efficiently for reasonably long times. The best result in this direction, due to Osborne \cite{Osborne}, shows that that time evolutions in one spatial dimension can be simulated efficiently (that is, with polynomial resources in $N$) only up to small times $t \approx \log(N)$. Instead, numerics suggest that the simulation of (\ref{eq:evolution}) can be performed efficiently for times that scale at least as the size of the system, $t \approx N$, \cite{TEBD2,tDMRG}. 

Several authors (see for instance \cite{TEBD2,White,Calabrese,Eisler,SpinCharge,Giorgi08}) have already considered particular instances of evolution ensuing either a \emph{local perturbation} of the ground state as in Eq. (\ref{eq:evolution}), or a \emph{local quench} (which is qualitatively equivalent to a local perturbation). In these specific cases it has been observed that, indeed, the amount of entanglement remains sufficiently small as to allow practical simulations for large times and system sizes (in a sense further specified in the next section). The goal of the present paper is to establish how general this result is, by conducting a systematic numerical study of evolutions of the form (\ref{eq:evolution}) in one-dimensional models, including: ($i$) systems of spins, fermions and bosons; ($ii$) gapped and gapless phases; ($iii$) homogeneous and disordered interactions; and ($iv$) integrable and non-integrable systems. We find that, whereas for each model the evolution may represent a very different physical process, in all cases the dynamics of entanglement, remarkably similar, are characterized by a moderate growth. As a result, the evolution can be simulated, with reasonably modest computational resources, for times long enough that the perturbation, initially localized, can spread over large regions, involving e.g. of the order of one hundred sites.

We notice that the TEBD algorithm has also been applied to other classes of evolution. In the case of a \emph{global quench} that modifies the Hamiltonian everywhere in the chain, or when the initial state of the system does belong to the low energy sector of $H$, it has been reported that a large amount of entanglement is created \cite{HalfChain,Disorder1} and therefore the simulation becomes inefficient. We emphasize that this is not at all surprising, given the exponentially large dimension of the Hilbert space in the system size, which implies that a generic state of the system cannot be represented efficiently. Instead, what is truly remarkable (and useful for practical simulations) is that there are circumstances of interest where the state of the system can be efficiently encoded using an MPS. Apart from the case of a local perturbation of the ground state/local quench, there are not many other classes of evolution known that can be efficiently simulated for reasonably large times. One noticeable exception is the case of a global quench in a \emph{disordered system} \cite{Disorder1}, which seems to be related to localization effects experienced by information in the presence of disorder \cite{Disorder2}. Another case, perhaps less surprising but still potentially very useful, is the evolution of observables and of thermal states after a global quench in \emph{integrable} systems \cite{Integrable}.

The rest of the paper is organized as follows. In Section II we introduce the measure of entanglement that is relevant in the present context and describe the one-dimensional models we have considered. In Section III we present the results of the simulations. Specifically, we describe two cases that illustrate the two types of scaling of entanglement observed in all the simulations we have performed. In Section IV we discuss the previous results, including simple toy models that reproduce the entanglement scaling reported in Section III, and present some conclusions.

\section{Entanglement Measure and one-dimensional models}

Our goal is to characterize the growth of entanglement during the time evolution that takes place after the ground state $\ket{\PsiGS}$ of a 1D system has been perturbed by some local operator $A$,  Eq. (\ref{eq:evolution}). In the present context, an appropriate measure of the entanglement contained in $\ket{\Psi(t)}$ is provided by the number $\chi$ of terms in its Schmidt decompositions (we refer to \cite{TEBD1,TEBD2} for details). More specifically, given a partition of the chain into two blocks containing the first $r$ sites and the remaining $N-r$ sites respectively, the Schmidt decomposition reads
\begin{equation}
	\ket{\Psi(t)} = \sum_{\alpha=1}^{\chi^r} \lambda^r_{\alpha} \ket{\Phi^{1:r}_{\alpha}}\ket{\Phi^{r+1:N}_{\alpha}},
\label{eq:Schmidt}
\end{equation}
where the rank $\chi^r$, the Schmidt coefficients $\lambda^r_{\alpha}$ and the Schmidt bases $\ket{\Phi^{1:r}_{\alpha}}$ and $\ket{\Phi^{r+1:N}_{\alpha}}$ are all time-dependent. Eq.\ (\ref{eq:Schmidt}) contains in principle a very large number of terms, with $\chi^r \sim \exp(N)$. However, when the Schmidt coefficients $\lambda_{\alpha}^r$ decay fast with $\alpha$, a good approximation to $\ket{\Psi(t)}$ may be obtained by keeping only a relatively small number $\chi_{\epsilon}^r$ of terms. The truncation introduces a small error $\epsilon$ \cite{truncationerror}, given by
\begin{equation}
	\epsilon \equiv 1 - \sum_{\alpha=1}^{\chi^r_{\epsilon}} (\lambda^r_{\alpha})^2.
\end{equation}
The exact and approximate ranks $\chi$ and $\chi_{\epsilon}$ of $\ket{\Psi(t)}$ are then defined by maximizing over bipartitions,
\begin{equation}
	\chi \equiv \max_{r}~ \chi^r,~~~~~~~~~~~\chi_{\epsilon}\equiv \max_{r}~ \chi_{\epsilon}^r.
\end{equation}
A MPS can store the truncated state using $O(N\chi_{\epsilon}^2)$ coefficients, while the cost of simulating a time step scales as $O(N\chi_{\epsilon}^3)$. Therefore $\chi_{\epsilon}$ is indicative of the computational resources required in an approximate simulation with truncation error $\epsilon$. Although in an actual simulation there might be other sources of errors, such as those due to Trotter expansion \cite{TEBD2}, and errors accumulate in time, the simple entanglement measure $\chi_{\epsilon}$ turns out to be informative enough as to allow us to assess in practice whether a simulation is efficient.

We have considered a large number of quantum models on a one dimensional lattice. Our hope is that by studying these models, we already observe all possible forms of scaling of entanglement, as measured by $\chi_{\epsilon}$, so that we can draw conclusions that are valid for generic 1D systems ---or at least for those 1D systems that are usually of interest in condensed matter physics and quantum statistical mechanics. The specific models we have considered are:

\noindent\textbf{1.---}The \emph{quantum Ising model with parallel and transverse magnetic fields}
\begin{equation}
H_{\mbox{\tiny{Ising}}} = \sum_r \sigma_x^r \sigma_x^{r+1} + \sum_r \left( h_x \sigma_x^r + h_z \sigma_z^r \right),
\label{eq:Ising}
\end{equation}
where $h_x$ and $h_y$ are the intensity of uniform magnetic fields in the $x$ (parallel) and $z$ (perpendicular) directions. For $h_x=0$, a Jordan-Wigner transformation maps this quantum spin model (with spin $\frac{1}{2}$) into a model of free spinless fermions.

\noindent\textbf{2.---}\emph{The quantum XY model with transverse magnetic field}
\begin{equation}
H_{\mbox{\tiny XY}} = \sum_r \left( \frac{1\!+\!\gamma}{2} \sigma_x^r \sigma_x^{r+1} +  \frac{1\!-\!\gamma}{2} \sigma_y^r \sigma_y^{r+1}  +  h^r_z \sigma_z^r \right),
\label{eq:XY}
\end{equation}
where $\gamma$ is the anisotropy parameter and $h^r_z$ the intensity of a (possibly site-dependent) transverse magnetic field. For $\gamma=1$ we recover the quantum Ising model with transverse magnetic field, whereas $\gamma=0$ corresponds to the quantum XX model, which is also used as a model of hard-core bosons. Again, a Jordan-Wigner transformation maps this quantum spin model (with spin $\frac{1}{2}$) into a model of free spinless fermions.

\noindent\textbf{3.---}\emph{The quantum XXZ model with magnetic field in the z direction}
\begin{equation}
	H_{\mbox{\tiny XXZ}}\! = \sum_r \!\left( \sigma_x^r \sigma_x^{r+1} \!+ \sigma_y^r \sigma_y^{r+1} + \Delta \sigma_z^r \sigma_z^{r+1}  +  h^r_z \sigma_z^r \right),
\label{eq:XXZ}
\end{equation}
where $\Delta$ is the anisotropy parameter and $h^r_z$ the intensity of a (possibly site dependent magnetic field in the $z$ direction. For $\Delta=0$ we recover the $XX$ model, for $\Delta = 1$ a spin-$\frac{1}{2}$ Heisenberg quantum antiferromagnet and for $\Delta=-1$ a model locally equivalent to a spin-$\frac{1}{2}$ Heisenberg quantum ferromagnet. Through a Jordan-Wigner transformation, $H_{\mbox{\tiny XXZ}}$ becomes a model of interacting spinless fermions.

\noindent\textbf{4.---}\emph{The Bose-Hubbard model}
\begin{equation}
H_{\mbox{\tiny{BH}}} =\sum_{s} \left[ -J(b^{\dagger s}  b^{s+1}  + hc.) -  \mu n^s + U n^s(n^s - 1)\right],
\label{eq:BH} 
\end{equation}
a model of interacting bosons where $J$ is the hopping amplitude, $\mu$ is the chemical potential, $U$ is the on-site repulsion; and $b^{\dagger s}$, $b^s$ and $n^s=b^{\dagger s}b^s$ are the bosonic creation, annihilation and number operators at site $s$, respectively.

For all the above models we have considered gapped and gapless cases. For the XY and XXZ models we have considered, in addition to uniformed magnetic fields, disordered versions by introducing inhomogeneous, random magnetic fields $h^r_z$. 

\section{Results}

We have simulated time evolutions of the form (\ref{eq:evolution}) for each of the variants of the models (\ref{eq:Ising})-(\ref{eq:BH}) of the previous section, for chains with open boundary conditions and with a number $N$ ranging from 100 to 200. After computing the ground state $\ket{\PsiGS}$ of a given Hamiltonian $H$ through a simulation of imaginary time evolution \cite{TEBD2}, at $t=0$ an operator $A$ is applied to the middle of the chain. The system is then allowed to evolve according to the time evolution operator $e^{-iHt}$, that results in a propagation of the perturbation from the center of the chain toward its ends. We stop the simulations before the signal reaches the boundary of the chain. Chains of different lengths have been simulated in order to guarantee that our results are essentially independent of the system's size. Convergence of the results in this sense is only possible in those models where the correlation length is sufficiently smaller than the system's size. We notice that systems at or very close to a quantum critical point fail to fulfill this condition. Also, the simulations have been performed with an MPS of rank $\chi^*$ much larger than the reported $\chi_{\epsilon}$, in order to guarantee that our analysis is independent of the computational resources devoted to the numerical simulation used to obtain them. Specifically, we report results for $\chi_{\epsilon}\leq 35$ which have been obtained by simulating the evolution with an MPS with $\chi^* = 80,100$ and $150$.

The evolution in all simulations has been carried on with a fourth order Trotter expansion (labeled as ${\cal Z}^1_4$ in \cite{Sornborger99}) with a time step $\delta t =0.02$. The associated Trotter error is $\epsilon_{\mbox{\tiny Trott}} \propto \delta t^5 \times T/\delta t =  8\times10^{-6}$ for $T=50$ time units. The computation times varied from few hours to three weeks in the worst case, in a PC with dual core processor at 2.2GHz with 4GB of
RAM. In the Bose-Hubbard model, each site was truncated to four levels.

We observe that, in spite of the rather diverse nature of the physics described by the models (\ref{eq:Ising})-(\ref{eq:BH}), the scaling of its entanglement in time follows a very similar pattern. Next we describe in detail the results obtain for two specific  systems. They have been chosen as representatives of the behavior of $\chi_{\epsilon}$ in all the models under study.

\begin{figure}
\includegraphics[width=8.7cm]{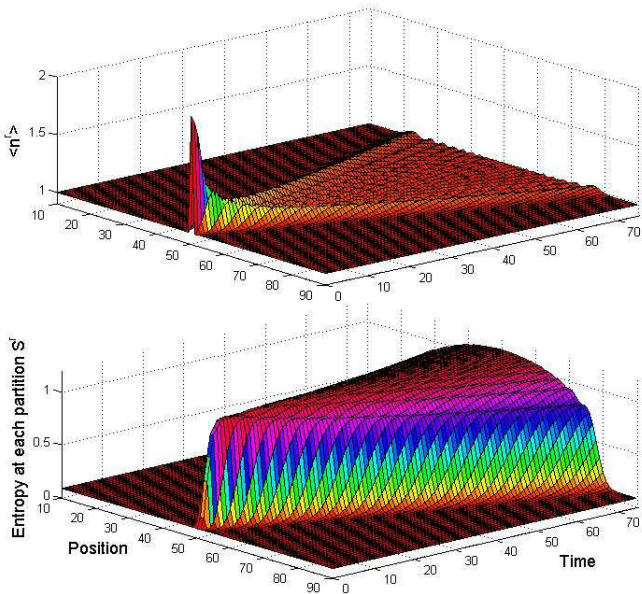}\\
\caption{
Propagation of a perturbation introduced in the middle of a chain of $N=100$ sites. This evolution corresponds to a Bose-Hubbard model, Eq.\ (\ref{eq:BH}), in the Mott-insulating phase (hopping term $J=0.1$, on-site repulsion $U=1$ and chemical potential $\mu=0.5$) with one boson per site. \textbf{(a)} Expectation value $\left<n^r\right>$ for the number operator on each site, as a function of time, after an extra boson is introduced on site $s_0=50$. The perturbation propagates throughout the chain, occupying a region that grows linearly in time until reaching the boundaries. \textbf{(b)} The entanglement entropy $S^r$ shows how every bipartition remains almost disentangled until the arrival of the wave front. When the front is gone far away, the entropy saturates.}
\label{HBMottnument3D}
\end{figure}

\begin{figure}
\includegraphics[width=8.7cm]{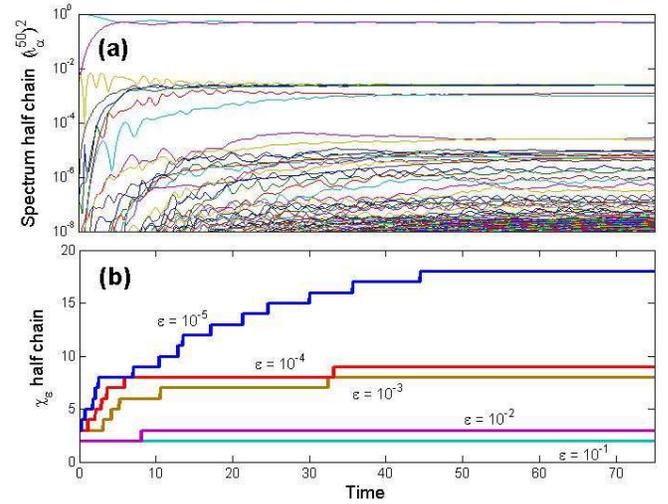}\\
\caption{Entanglement growth at the center of the chain for the Bose-Hubbard model, Eq.\ (\ref{eq:BH}), in the same scenario as Fig.\ (\ref{HBMottnument3D}). \textbf{(a)} Starting from a barely entangled state, with only a few non-vanishing Schmidt coefficients, several other coefficients start to grow, then saturate. \textbf{(b)} The approximate rank $\chi_{\epsilon}$ first grows, then saturates at a number that increases as we decrease the truncation error $\epsilon$. This implies that an accurate approximation to $\ket{\Psi(t)}$ can be stored in a MPS with a small rank $\chi_{\epsilon}$ that after some time becomes roughly constant.}
\label{HBMottspectrumchieps100}
\end{figure}

\textbf{Example I: Saturation.---} Our first example is concerned with a system of interacting bosons described by the Bose-Hubbard model $H_{\mbox{\tiny{BH}}}$ of Eq. (\ref{eq:BH}) in a chain with $N=100$ sites. Fig.\ (\ref{HBMottnument3D}a) shows the reorganization of the density of bosons when one extra particle is introduced in the middle of the chain, which is in the Mott insulating phase. Fig.\ (\ref{HBMottspectrumchieps100}a) presents the changes in the spectrum of the squared Schmidt coefficients $(\lambda^{r=50}_{\alpha})^2$ during the evolution. We depict the Schmidt coefficients for $r=50$ (middle of the chain) because this is the bipartition that appears most entangled (i.e., $\chi_{\epsilon} = \chi_{\epsilon}^{r=50}$ for all $\epsilon$) at all times (except some minor corrections of one or two sites). We measure time in $1/U$ units, and choose $U=1$.

We notice that there are two marked regimes. First, for times up to  $t=20-30$, a number of Schmidt coefficients increase monotonically, with a growth that progressively slows down. Bipartite entanglement between the left and the right halves of the chain is being created (see an analogous behavior obtained analyticaly in Ref. \cite{Giorgi08}). Then, for larger times, and coinciding with the fact that the fronts of the density wave are far from the middle of the chain, the Schmidt coefficients become roughly stationary, indicating that the production of bipartite entanglement at the center of the chain has come to a halt. Notice that this is reflected in an initial growth, then saturation, of $\chi_{\epsilon}$, as depicted in Fig.\ (\ref{HBMottspectrumchieps100}b). Elsewhere in the chain, say $r=30$, no bipartite entanglement is generated until the front of the density wave arrives. At that point, the Schmidt coefficients $\lambda^{30}_{\alpha}$ start increasing to later become stationary, in a pattern that mimics what occurred at the center of the chain, but with increasingly delayed and attenuated growth as we move away from the center.
The entanglement entropy of a bipartition,
\begin{equation}
	S^r \equiv - \sum_{\alpha} (\lambda_{\alpha}^r)^2\log_2 (\lambda_{\alpha}^r)^2,
\end{equation}
offers a complementary, coarse-grained picture into the entanglement growth in the system, that confirms the above observations, see Fig.\ (\ref{HBMottnument3D}b).

\begin{figure}
\includegraphics[width=8.6cm]{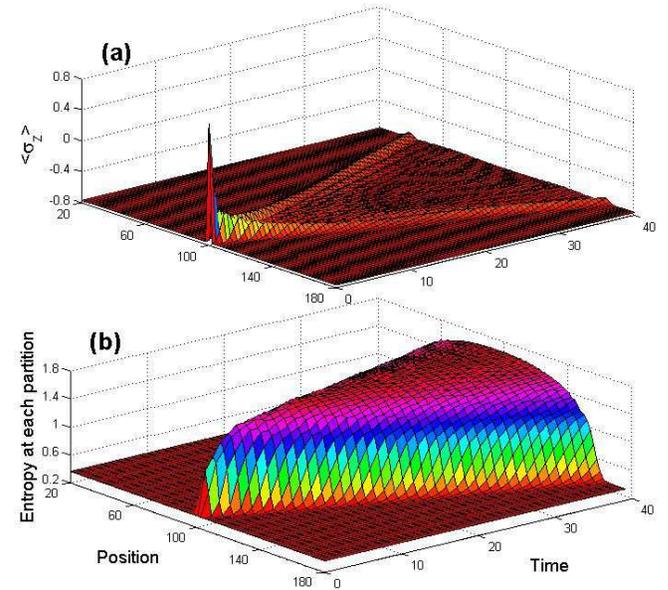}\\
\caption{Propagation of a perturbation introduced in the middle of a chain with $N=200$ sites, corresponding to the Ising model with tilted magnetic field, Eq.\ (\ref{eq:Ising}), with $h_x=h_z=1$. Starting with a slightly entangled ground state, at $t=0$ one spin is flipped on site 100 applying $\sigma_x^{r=100}$. The evolution of both \textbf{(a)} the expectation value $\left<\sigma_z^r\right>$ for the magnetization on the $z$ direction and \textbf{(b)} the entropy of a bipartition resemble their analogues in the Bose-Hubbard model, Fig.\ (\ref{HBMottnument3D}). However, Fig.\ (\ref{IsBtiltAF200spectrum150chieps80-100-150}) reveals that more entanglement is generated in this second system.}
\label{IsBtiltAF200entsgz3Dchi150}
\end{figure}

\begin{figure}
\includegraphics[width=8.7cm] {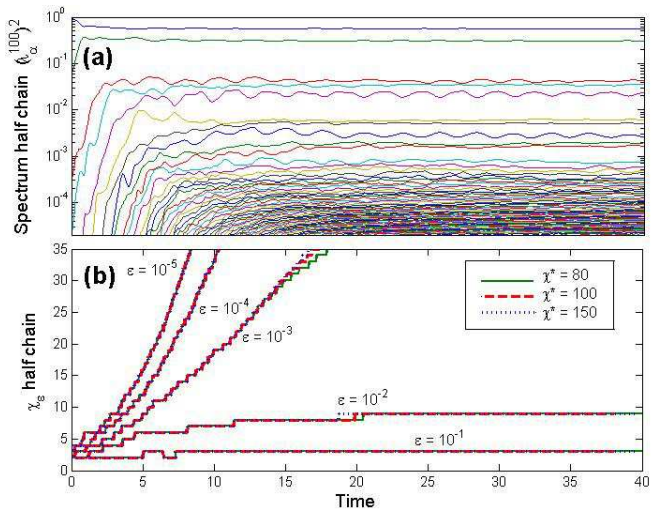}\\
\caption{Entanglement growth at the center of the chain for the Ising model with tilted magnetic field, Eq.\ (\ref{eq:Ising}) in the same scenario as Fig.\ (\ref{IsBtiltAF200entsgz3Dchi150}). 
\textbf{(a)} As in Fig. (\ref{HBMottspectrumchieps100}a), starting from a barely entangled state, with only a few non-vanishing Schmidt coefficients, several other coefficients start to grow, then tend to saturate. \textbf{(b)} But while rank $\chi_\epsilon$ still saturates for $\epsilon=0.1,0.01$, it keeps growing (approximately linearly) for lower values $\epsilon=10^{-3},10^{-4}$. [Notice that we have superposed the results of simulations performed with an MPS of increasing rank $\chi^*=80,100,150$. The results are generally very similar (except e.g. for a small variation in $\chi_{\epsilon=10^{-3}}$)].}
\label{IsBtiltAF200spectrum150chieps80-100-150}
\end{figure}

\textbf{Example II: Moderate steady growth.---} A slightly different pattern of entanglement growth occurs in the antiferromagnetic quantum Ising model with tilted magnetic field $H_{\mbox{\tiny{Ising}}}$ of Eq. (\ref{eq:Ising}) with $N=200$ spins. Fig.\ (\ref{IsBtiltAF200entsgz3Dchi150}a) shows the propagation of a magnetization wave produced by flipping one spin in the center of the chain, $r=100$, for the case $h_x=h_z=1$. Fig.\ (\ref{IsBtiltAF200spectrum150chieps80-100-150}a) shows that, as in the first example, a number of Schmidt coefficients first grow substantially and then tend to saturate. An important difference, however, is revealed by studying $\chi_{\epsilon}$, Fig.\ (\ref{IsBtiltAF200spectrum150chieps80-100-150}b). For $\epsilon = 0.1, 0.01$, the approximate rank $\chi_{\epsilon}$ again saturates, but for smaller $\epsilon$ it keeps growing in time. This is due to the appearance of an increasingly large number of small Schmidt coefficients, that need to be consider collectively in order to decrease the truncation error $\epsilon$. Although it is difficult to characterize the scaling of $\chi_{\epsilon}$ in time, we obtain results compatible with $\chi_{\epsilon}\sim t^p$ for a power $p$ of the order of 1. Correspondingly, a plot of the entropy, Fig.\ (\ref{IsBtiltAF200entsgz3Dchi150}b), shows an initial rapid growth followed by a much slower, possibly logarithmic growth at longer times.

\section{Discussion}

Our simulations show a strikingly similar pattern of entanglement generation in all the systems we have studied \cite{critical}. In particular, we always find a sufficiently large $\epsilon$, e.g. $\epsilon = 10^{-2}$ in the examples of the previous section, for which $\chi_{\epsilon}$ seems to saturate as a function of time. This means that an MPS with fixed rank $\chi_{\epsilon}$ already suffices to approximate (to accuracy $\epsilon$) the state of the chain during the \emph{whole} evolution. However, if we now try to improve this accuracy by lowering $\epsilon$, we find two possible behaviors. In some cases, as in example I, the rank $\chi_{\epsilon=10^{-3}}$ still saturates, although at a later time. In some other cases, as in example II, $\chi_{\epsilon=10^{-3}}$ keeps growing indefinitely (that is, as far as we could see with our computational resources), with a shape that resembles a small power of time $\chi_{\epsilon}\sim t^p$. 
This moderate entanglement growth agrees with the analytical result obtained in Ref.\ \cite{Calabrese} in a similar scenario. By means of conformal field theory they show that the entropy grows logarithmically with time, $S(t) \sim  \log(t)$, when two initially decoupled quantum chains are joined together. This corresponds as well to a local quench of the ground state, and the subsequent evolution is expected to be analogous in terms of entanglement creation through the whole chain. 
These mild behaviors are in sharp contrast with the exponential growth, $\chi_{\epsilon}\sim e^t$, that one finds e.g., in a spin model where at time $t=0$ the value of the magnetic field throughout the whole chain is changed (global quench) or the spins in half of the chain are flipped \cite{Calabrese,HalfChain,Disorder1}.

Although the scaling of $\chi_{\epsilon}$ described in this paper corresponds to \emph{large} times in a \emph{large} systems, both the time and the system size are obviously finite. Our results show that at least during a valuably large period of time, the entanglement grows in a moderate way, enabling the efficient, accurate simulation of the evolution. Whether at even larger times the simulation remains efficient is hard to tell. This is not too relevant, however, in all those applications where the interesting phenomena occurs within moderate times from the moment the perturbation is introduced. 

In conclusion, we have studied the generation of entanglement in a particularly relevant class of time evolutions, namely those that follow from locally perturbing the ground state of a system, which are related to the computation of unequal-time two-point correlators (\ref{eq:correlator}) in one spatial dimension. Our results provide strong evidence that such evolutions can quite often be efficiently simulated with algorithms, such as the TEBD, based on updating a MPS \cite{TEBD1,TEBD2,TEBD3,White,Daley,tDMRG}. Our expectation is that the scaling of entanglement that we have observed for the specific models under consideration also applies to most one-dimensional models of interest in condensed matter physics and quantum statistical mechanics. It would be extremely interesting to be able to formally justify these results. This does not seem to be an easy task. 

\begin{figure}
\includegraphics[width=5cm]{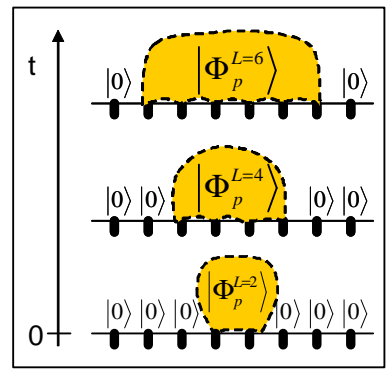}\\
\caption{
Toy model for the time evolutions of examples I and II of Sect. III. The chain is initially in a ground state with all spins in state $\ket{0}$. Either 1 or 2 spins are flipped into state $\ket{1}$ at the middle of the chain, and the evolution produces an entangled state $\ket{\Phi^L_p}$ of Eqs. (\ref{eq:Phi1}) and (\ref{eq:Phi2}) ($p=1$ or $2$) that involves a number $L$ of central spins that grows linearly in time.}
\label{fig:ToyModels}
\end{figure}

We conclude by describing two simple toy model evolutions that reproduce the two observed patterns of entanglement scaling. Initially, a spin chain is in a ground state where all the spins are in state $\ket{0}$. In one case, the perturbation flips one single spin into state $\ket{1}$, and the time evolution produces an entangled state
\begin{equation}
\ket{\Phi_1^L} \equiv \frac{1}{\sqrt{L}} \sum_{i} \ket{0_1\cdots 1_{i}\cdots 0_L}
\label{eq:Phi1}
\end{equation}
involving the $L$ central spins of the chain, where $L$ grows linearly in time, $L=2vt$, with $v$ is the speed of propagation of the perturbation, see Fig. (\ref{fig:ToyModels}). Here, $\ket{\Phi^L_1}$ is a linear combination of all the strings of $L$ spins containing one single one. It can be seen that in any link of the chain, the Schmidt rank $\chi$ will go from 1 (before the expanding perturbation has reached that link) to just 2. Thus, this model reproduces the saturation observed in the example I.
In the second case, the perturbation flips two spins into state $\ket{1}$ and the time evolution produces an entangled state
\begin{equation}
\ket{\Phi_2^L} \equiv \frac{1}{\sqrt{L(L-1)}} \sum_{i,j\neq i} \ket{0_1\cdots 1_{i} \cdots 1_{j}\cdots 0_L}
\label{eq:Phi2}
\end{equation}
involving the $L$ central spins of the chain, where as before $L$ grows linearly in time, $L=2vt$. Now $\ket{\Phi^L_2}$ is a linear combination of all the strings of $L$ spins containing two ones. It can be seen that in any link of the chain, the Schmidt rank $\chi$ grow linearly in time from the moment the expanding perturbation reaches that link. Therefore this second evolution reproduces the moderate steady growth observed in the example II. Perhaps then (more sophisticated versions of) these toy models will inspire new simulation algorithms, based on an alternative ansatz that is even more efficient than an MPS for the considered class of time evolution.

 
A.P.\ thanks A. Doherty for his hospitality at University of Queensland,  and acknowledges financial support from Spanish MEC (Programa para la Movilidad) and from Universidad de Alcal\'a (Ayudas a la Movilidad). G.V.\ acknowledges financial support from Australian Research Council (FF0668731, DP0878830).


\begin{thebibliography}{}
\bibitem{truncationerror} The truncation error refers to the whole state of the chain. The induced (relative) errors in the expectation value of a local observables are often up to several orders of magnitude smaller than $\epsilon$.
\bibitem{critical} As a (finite size) model approaches a quantum critical point, its ground state displays more and more entanglement. Consequently, a larger $\chi_{\epsilon}$ is required in order to achieve the same accuracy $\epsilon$. However, the scaling of $\chi_{\epsilon}$ with time still seems to follow the same patterns as in the non-critical case.
\end{thebibliography}
\end{document}